\documentstyle[12pt]{article}
\begin{document}
\thispagestyle{empty}
\begin{center}
\LARGE\tt\bf{Ghost neutrinos and radiative Kerr metric in Einstein-Cartan gravity}
\end{center}
\vspace{2.5cm}
\begin{center}{\large L.C. Garcia de Andrade\footnote{Departamento de F\'{\i}sica Teorica-UERJ.
Rua S\~{a}o Fco. Xavier 524, Rio de Janeiro, RJ
Maracan\~{a}, CEP:20550-003 , Brasil. E-Mail.: garcia@dft.if.uerj.br}}
\end{center}
\vspace{2.0cm}
\begin{abstract}
Ghost neutrino solution in radiative Kerr spacetime endowed with totally skew-symmetric Cartan contortion is presented. The computations are made by using the Newman-Penrose (NP) calculus. The model discussed here maybe useful in several astrophysical applications specially in black hole astrophysics.
\end{abstract}
\vspace{2.0cm}
\vspace{0.5cm}
\noindent
PACS number(s): 0420, 0450
\newpage

Earlier Audretsch \cite{1} have presented two interesting types of ghost neutrino solutions in Riemannian spacetime having as its carachteristic the fact that the the energy-momentum tensor was null while the neutrino current vector did not vanish. In this paper he showed that the Scharzschild and Kerr-Newman metric were not able to support ghost neutrinos. Besides he presented a solution of the Weyl neutrino equation in General Relativity (GR) which represented a pp gravitational wave coupled with a ghost neutrino. More recently Griffiths \cite{2} presented a ghost neutrino solution in Einstein-Cartan (EC) Weyl equation which generalized the Collinson-Morris \cite{3} ghost neutrino in GR. In the present letter we show that is possible to have a new solution of ECWeyl equation representing a ghost neutrino in Kerr radiative spacetime with torsion. Despite of several ghostness conditions proposed earlier by Letelier \cite{4} we addopt here the same one as given in Griffiths which is the vanishing of the Riemann-Cartan $U_{4}$ Ricci tensor where 
\begin{equation}
R_{{\mu}{\nu}}({\Gamma})=0
\label{1}
\end{equation}
where ${\Gamma}$ is the $U_{4}$ connection $({\mu}=0,1,2,3)$. The $J^{\mu}$ is the neutrino current vector given by
\begin{equation}
J^{\theta}={\phi}(u)l^{\theta} 
\label{2}
\end{equation}
where ${\phi}$ is the neutrino field. The Ricci tensor in $U_{4}$ can be expressed in terms of the Ricci tensor in the Riemannian manifold $V_{4}$ as 
\begin{equation}
R_{({\mu}{\nu})}({\Gamma})= {R^{0}}_{{\mu}{\nu}}+{\nabla}_{\alpha}{K_{{\mu}{\nu}}}^{\alpha}-{K_{{\alpha}{\mu}}}^{\beta}{K_{{\beta}{\nu}}}^{\alpha}  
\label{3}
\end{equation}
where the round brackets indicate the symmetrization and the zero superscript indicates the Riemannian quantities. Thus the symmetric part of the Ricci-Cartan tensor is given by 
\begin{equation}
R_{({\mu}{\nu})}({\Gamma})= {R^{0}}_{{\mu}{\nu}}+2k^{2} J_{{\mu}}J_{{\nu}}  
\label{4}
\end{equation}
which can be expressed in therms of the neutrino current scalar field ${\phi}$ as
\begin{equation}
R_{({\mu}{\nu})}({\Gamma})= {R^{0}}_{{\mu}{\nu}}+2k^{2}{\phi}^{2}(u) l_{{\mu}}l_{{\nu}}  
\label{5}
\end{equation}
Here the vector $l^{\mu}$ represents one of the four legs of the tetrad of null vectors defined by
\begin{equation}
e^{\mu}_{i}= (l^{\mu}, n^{\mu},m^{\mu},\bar{m^{\mu}})
\label{6}
\end{equation}
where $i=1,2,3,4$ and $l^{\mu}$ and $n^{\mu}$ are the real vectors and $m^{\mu}$ and $\bar{m^{\mu}}$ are complex conjugate. The tetrad indices i are lowered and raised by the tetrad Minkowski metric ${\eta}_{mn}$ which only nonvanishing components are ${\eta}_{01}=1$ and ${\eta}_{23}= -1$. Now let us apply this equation to the Kerr radiative metric \cite{5} in the coordinates $x^{0}=u$, $x^{1}=r$,$x^{2}=x$ and $x^{3}=y$ where the line element is given by
\begin{equation}
ds^{2}= (1-2mr{\rho}\bar{\rho})du^{2}+2dudr+4mrasin^{2}x{\rho}\bar{\rho}dudy -2asin^{2}xdrdy-{({\rho}\bar{\rho})}^{-1}dx^{2}-fdy^{2}
\label{7}
\end{equation}
where
\begin{equation}
f:= 2mra^{2}sin^{2}x {\rho}\bar{\rho}+ r^{2}+a^{2}sin^{2}x
\label{8}
\end{equation}
here $u=t-r$ is the retarded time coordinate and the speed of light in vacuum $c=1$. The expression $m(u)$ is the mass parameter. Besides a is a constant parameter like in the Kerr \cite{6} metric. where we have used the geometrical optics approximation up to order of $O(r^{-2})$ and drop out terms of order $O(r^{-3})$. This approximation allows us to obtain a very simple ghost neutrino solution. The spin coefficient ${\rho}$ appearing in the line element is given by 
\begin{equation}
{\rho}= -\frac{1}{(r-iacosx)}
\label{9}
\end{equation}
The null tetrad for this metric is computed by making use of the variables
\begin{equation}
{\Omega}=r^{2}+a^{2}
\label{10}
\end{equation}
and 
\begin{equation}
Y=\frac{r^{2}+a^{2}-2m(u)r}{2}
\label{11}
\end{equation}
The null tetrad for this metric is computed by making use of the variables
\begin{equation}
l_{\mu}={\delta}_{\mu}^{0}-a sin^{2}x {\delta}_{\mu}^{3}
\label{12}
\end{equation}
\begin{equation}
m_{\mu}= -\frac{\bar{\rho}}{\sqrt{2}}(iasinx{\delta}^{0}_{\mu}-({\rho}\bar{\rho})^{-1}{\delta}^{2}_{\mu}-i{\Omega}sinx{\delta}^{3}_{\mu})
\label{13}
\end{equation}
\begin{equation}
n_{\mu}= {\rho}\bar{\rho}[Y{\delta}_{\mu}^{0}-({\rho}\bar{\rho})^{-1}{\delta}^{1}_{\mu}-asin^{2}x{\delta}_{\mu}^{3}Y]
\label{14}
\end{equation}
From this tetrad one is able to show that the following spin-coefficients vanish
\begin{equation}
{\epsilon}^{0}={\lambda}^{0}={\sigma}^{0}={\kappa}^{0}=0
\label{15}
\end{equation}
and 
\begin{equation}
{\pi}= \frac{iasinx {\rho}^{2}}{\sqrt{2}}
\label{16}
\end{equation}
\begin{equation}
{\beta}= -\frac{cotx \bar{\rho}}{2\sqrt{2}}
\label{17}
\end{equation}
\begin{equation}
{\alpha}= {\pi}-\bar{\beta}
\label{18}
\end{equation}
\begin{equation}
{\mu}= Y{\rho}^{2}\bar{\rho}
\label{19}
\end{equation}
\begin{equation}
{\nu}=-i{\dot{\bar{m}}}ra \frac{sinx {\rho}^{2}\bar{\rho}}{\sqrt{2}}
\label{20}
\end{equation}
\begin{equation}
{\gamma}= {\mu}+[r-m(u)]\frac{{\rho}\bar{\rho}}{\sqrt{2}}
\label{21}
\end{equation}
\begin{equation}
{\tau}=-ia \frac{sinx {\rho}^{2}\bar{\rho}}{\sqrt{2}}
\label{22}
\end{equation}
The radiative Kerr metric reduces to the Vaidya metric when the angular momentum of the compact object, a black hole or very massive star, vanishes. Now let us substitute the Riemannian Ricci tensor 
\begin{equation}
{R^{0}}_{{\mu}{\nu}}=- 2\dot{m}(u)r^{2}{({\rho}\bar{\rho})}^{2}l_{\mu}l_{\nu}
\label{23}
\end{equation}
Substitution of expression (\ref{22}) into expression (\ref{5}) yields
\begin{equation}
\dot{m}(u)=k^{2}\frac{{\phi}^{2}(u)}{r^{2}{({\rho}\bar{\rho})}^{2}} 
\label{24}
\end{equation}
where we have applied the ghostness condition in ECWeyl spacetime $(R_{{\mu}{\nu}}(\Gamma)=0)$ to obtain this relation between the mass loss parameter $m(u)$ and the neutrino current scalar field ${\phi}(u)$. Let us now assume that the neutrino current is constant which implies that the function ${\phi}(u)={\phi}_{0}=constant$. Thus from expression (\ref{23}) one obtains the following simple solution
\begin{equation}
{m}(u)=k^{2}\frac{{{\phi}^{2}}_{0}u}{r^{2}{({\rho}\bar{\rho})}^{2}} 
\label{25}
\end{equation}
Substitution of this solution into the line element (\ref{8})
\begin{equation}
ds^{2}= (1-2k^{2}\frac{{{\phi}_{0}}^{2}u}{r{\rho}\bar{\rho}})du^{2}+2dudr+[4k^{2}\frac{{{\phi}_{0}}^{2}u}{{\rho}\bar{\rho}}du -2adr]sin^{2}xdy-{({\rho}\bar{\rho})}^{-1}dx^{2}-fdy^{2}
\label{26}
\end{equation}
where
\begin{equation}
f:= [(a^{2}+2k^{2}\frac{{{\phi}_{0}}^{2}u}{r{\rho}\bar{\rho}})sin^{2}x+ r^{2}]
\label{27}
\end{equation}
\section*{Acknowledgements}

\paragraph*{}
I am very much indebt to Professor P.S.Letelier for helpful discussions on this subject of this paper. Thanks are also due to Professor Paul Tod for suggesting me to work with NP calculus with torsion many years ago during my post-doctoral at the Mathematical Institute of Oxford University. Grants from CNPq (Ministry of Science of Brazilian Government) and Universidade do Estado do Rio de Janeiro (UERJ) are acknowledged.

\end{document}